\documentstyle[12pt]{article}
\begin{document}
\title{SUSY QM VIA 2x2 MATRIX SUPERPOTENTIAL} 
\author{
R. de Lima Rodrigues\footnote{Permanent address: Departamento de
Ci\^encias Exatas e da Natureza, Universidade Federal da Para\'\i ba,
Cajazeiras - PB, 58.900-000 - Brazil. 
E-mail to RLR is rafaelr@cbpf.br or
rafael@fisica.ufpb.br and E-mail to ANV is vaidya@if.ufrj.br.}
\\ Centro Brasileiro de Pesquisas F\'\i sicas (CBPF)\\
Rua Dr. Xavier Sigaud, 150, CEP 22290-180\\ Rio de Janeiro, RJ, Brazil
and\\
A. N. Vaidya\\
Instituto de F\'\i sica\\
Universidade Federal do Rio de Janeiro\\
Caixa Postal 68528 - CEP 21945-970, Rio de Janeiro}

\date{}
\maketitle

\begin{abstract} 
   The $N=2$ supersymmetry  in quantum mechanics 
involving two-component eigenfunction is investigated.
\end{abstract}

\vspace{0.5cm}

\newpage
\section{Introduction}

\paragraph*{}

The algebraic technique of the supersymmetry in quantum mechanics (SUSY QM) 
formulated by Witten \cite{W}, in which the essential idea is 
based on the Darboux 
procedure on second-order differential equations, 
has been extended in order to find the 2x2 matrix superpotencial
 \cite{semenove91,ss2,Ta93,ger03}. 

  In this work we show that the superpotential for the SUSY QM 
with two-component 
wave functions is a Hermitian matrix, and we consider the
applicatoion to a planar physical system,
a neutron interacting with the magnetic field  \cite{RVV}.

\section{Supersymmetry for two-component eigenfunction}

\paragraph*{}

 In this section we consider a non-relativistic Hamiltonian ({\bf H}$_1$) 
for a two-component wave function in the following bilinear forms
 
\begin{eqnarray}
\label{E10}
\hbox{{\bf H}}_1 
&&= \hbox{{\bf A}}_1^+ \hbox{{\bf A}}_1^-+E_1^{(0)}\nonumber\\
{}&&=
-\hbox{{\bf I}}\frac{d^2}{dx^2} +
\left(\frac{d}{dx}\hbox{{\bf W}}_{1}(x)\right)+
\hbox{{\bf W}}_{1}(x)\frac{d}{dx}-
\hbox{{\bf W}}_{1}^{\dag}(x)\frac{d}{dx}
+\hbox{{\bf W}}_{1}^{\dag}\hbox{{\bf W}}_{1}(x),
\end{eqnarray}

\begin{eqnarray}
\label{E11}
\hbox{{\bf H}}_2 
&&= \hbox{{\bf A}}_1^- \hbox{{\bf A}}_1^++E_1^{(0)}\nonumber\\
&&=
-\hbox{{\bf I}}\frac{d^2}{dx^2} -
\left(\frac{d}{dx}\hbox{{\bf W}}_{1}^{\dag}(x)\right)
-\hbox{{\bf W}}_{1}^{\dag}(x)\frac{d}{dx}+\hbox{{\bf W}}_{1}(x)\frac{d}{dx}
+\hbox{{\bf W}}_{1}(x)\hbox{{\bf W}}_{1}^{\dag},
\end{eqnarray}
where
\begin{equation}
\label{EA}
\hbox{{\bf A}}_1^{-}=-\hbox{{\bf I}}\frac{d}{dx} +\hbox{{\bf W}}_{1}(x),
\quad \hbox{{\bf A}}_1^+=\left( \hbox{{\bf A}}_1^-\right)^{\dag}.
\end{equation}

So far $\hbox{{\bf W}}_{1}(x)$ can be a two by two non-Hermitian matrix, but 
we will now show that {\bf H}$_1$ and {\bf H}$_2$
are exactly the Hamiltonians of the bosonic and fermionic
sectors of a SUSY Hamiltonian if and only if the matrix superpotential is 
a Hermitian one. Indeed (comparing the pair SUSY Hamiltonians {\bf H}$_{\pm}$
with the Hamiltonians {\bf H}$_1$ and {\bf H}$_2$)
we see that only when the hermiticity condition of the 
{\bf W}$_{1}$ is readily satisfied, i.e.,
$\hbox{{\bf W}}_{1}^{\dag}=\hbox{{\bf W}}_{1},
$
we may put {\bf H}$_1$ in a bosonic sector Hamiltonian. 
In this case {\bf H}$_1$ ({\bf H}$_2$) becomes exactly 
{\bf H}$_-$ ({\bf H}$_+$) of a SUSY Hamiltonian model, 
analogous to the Witten model, viz.,

\begin{equation}
\label{E13a}
 {\bf H}_{SUSY} = -
{1\over 2} {\bf I}{d^{2}\over dx^{2}} + {1\over 2} \left \{{\bf W}^{2}(x) + 
{\bf W}^{\prime}(x) \sigma _{3}\right\} 
= \pmatrix{{\bf H}_{-}&0\cr0&{\bf H}_{+}},
\end{equation}
where $\sigma_3$ is the Pauli matrix. Only under the hermiticity condition
 one can
to call {\bf W}$_{1}={\bf W}(x)$ of a matrix superpotential.

Let {\bf H}$_{\pm}$ be the bosonic (-) and fermionic (+) sector Hamiltonians 
for a two-component eigenstate $\Psi_-$, given by
 
\begin{equation}
\label{E8a}
\hbox{{\bf H}}_{\pm} = 
-\hbox{{\bf I}}\frac{d^2}{dx^2} +\hbox{{\bf V}}_{\pm}(x), 
\quad \Psi_-(x)=\left(
\begin{array}{c}
\psi_{-,1}(x) \\ 
\psi_{-,2}(x)
\end{array}\right), \quad E_-^{(0)}=0,
\end{equation}
where {\bf I} denotes the 2x2 unit matrix and the pair of SUSY
potential {\bf V}$_-(x)$, is 
a 2x2 matrix potential which may be written in terms of
a 2x2 matrix superpotential {\bf W}(x), viz.,

\begin{equation}
\label{E9a}
\hbox{{\bf V}}_{\pm}(x) =
\hbox{{\bf W}}^2(x)\mp\hbox{{\bf W}}^{\prime}(x).
\end{equation}

Let us consider the eigenvalue equations for the bosonic and 
fermionic sector Hamiltonians, viz.,

\begin{equation}
\label{E14}
\hbox{{\bf H}}_{\pm} \Psi^{(n)}_{\pm}
= E^{(n)}_{\pm}\Psi^{(n)}_{\pm}, n=0, 1, 2, \cdots .
\end{equation}
These systems can exhibit bound and continuous eigenstates under the 
annihilation conditions 

\begin{equation}
\label{E15}
\hbox{{\bf A}}^-\Psi^{(0)}_-=0, \quad \Psi_-^{(0)}(x)=\left(
\begin{array}{c}
\psi_{-,1}^{(0)}(x) \\ 
\psi_{-,2}^{(0)}(x)
\end{array}\right)
\end{equation}
or
\begin{equation}
\label{E16}
\hbox{{\bf A}}^+\Psi^{(0)}_+=0, \quad \Psi_+^{(0)}(x)=\left(
\begin{array}{c}
\psi_{+,1}^{(0)}(x) \\ 
\psi_{+,2}^{(0)}(x)
\end{array}\right).
\end{equation}
In this case we see that one cannot put $\Psi_+^{(0)}(x)$ in terms
of $\Psi_-^{(0)}(x)$ and vice-versa in a similar manner to the case
of one-component eigenfunction system.
However, if $\Psi^{(0)}_-(x)$ is normalizable we have

\begin{equation}
\label{E17}
\int_{-\infty}^{+\infty}\left(\vert\psi^{(0)}_{-,1}\vert^2+
\vert\psi^{(0)}_{-,2}\vert^2\right)dx=1.
\end{equation}

Note that in Eq. (5) of ref. \cite{Ta93} 
the author has taken a particular Hermitian matrix for his superpotential 
in such a way that the validity of his development is ensured.

Let us now consider the interesting application of the above
development  for a 
bidimensional physical system 
in coordinate space associated to a Neutron with magnetic momentum 
$\vec\mu=\mu(\sigma_1, \sigma_2, \sigma_3)$ in a static magnetic field
\cite{RVV}. In this case, $x=\rho>0,$
the Ricatti equation in matrix form is given by

\begin{equation}
\mbox{\bf V}_-(\rho)=\mbox{\bf W}^{\prime}(\rho)+\mbox{\bf W}^2(\rho) =
\left(\begin{array}{llll} \frac{m^2-\frac{1}{4}}{\rho^2} &
\;\;\;\frac{-2F}{\rho}\\ \frac{-2F}{\rho} &
\frac{(m+1)^2-\frac{1}{4}}{\rho^2}
\end{array}
\right)-\mbox{\bf I}{\tilde
E}^{(0)}_1,
\end{equation}
which has the following particular solution for the 2x2 matrix superpotential
given by 
$
\mbox{\bf W}_m=\left(\begin{array}{ll}
\frac{m+\frac{1}{2}}{\rho}  &  -\frac{F}{m+1}\\
-\frac{F}{m+1}  & \frac{m+\frac{3}{2}}{\rho}
\end{array}
\right),
$
where the energy
eigenvalue of the ground state is
${\tilde E}^{(0)}_1=-\frac{F^2}{2(m+1)^2}, \quad
F\propto -\mu I, \quad m=0, \pm 1,
\pm 2, \cdots,$ and
 $\rho$  is the usual cylindrical coordinate.
We are considering the current $I$ located 
along the
$z$-axis, and we have used units
with $\hbar=1=mass.$ The current $I$ generate a static magnetic field.
Also, note that  $\mbox{\bf V}_-(\rho)$
has zero ground state energy, $E^{(0)}_{-} = 0,$ thus
SUSY is said to be unbroken.

The algebra of SUSY in quantum mechanics is characterized 
by one anti-commutation and two commutation
relations given below 

\begin{equation}  
\label{ass1}
H_{SUSY} = [Q_- ,Q_+ ]_+, \quad
\left[H_{SUSY}, Q_{\pm}\right]_- = 0 = (Q_-)^2 = (Q_+)^2.
\end{equation}
One representation of the $N=2$ SUSY superalgebra is the following

\begin{equation}  
\label{E40}
H_{SUSY} = [Q_- ,Q_+ ]_+ = \left( 
\begin{array}{cc}
\mbox{\bf A}^+ \mbox{\bf A}^- & 0 \\ 
0 & \mbox{\bf A}^-\mbox{\bf A}^+
\end{array}
\right)_{4\hbox{x}4}= \left( 
\begin{array}{cc}
{\bf H}_- & 0 \\ 
0 & {\bf H}_+
\end{array}
\right).
\end{equation}
The supercharges $Q_{\pm}$ are differential operators of first order and 
can be given by
$Q_- = \left( 
\begin{array}{cc}
0 & 0 \\ 
\mbox{\bf A}^- & 0
\end{array}
\right)_{4\hbox{x}4}, \quad 
Q_+ = \left( 
\begin{array}{cc}
0 & \mbox{\bf A}^+\\
0 & 0 
\end{array}
\right)_{4\hbox{x}4},$
where $\mbox{\bf A}^{\pm}$ are 2x2 non-Hermitian matrices given by 
Eq. (\ref{EA}).

\section{Conclusion}

\paragraph*{}

In this work we 
investigate an extension of the supersymmetry in non-relativistic
quantum mechanics for two-component wave functions.
This 
leads to 4x4 supercharges and supersymmetric Hamiltonians whose 
bosonic sectors are privileged with two-component eigenstates. 

We have considered the 
application for a Neutron in
interaction with a static magnetic field of a straight
current carrying wire 
\cite{RVV}.

\centerline{\bf ACKNOWLEDGMENTS}

This research was supported in part by CNPq (Brazilian Research Agency).
The authors are also grateful to the organizing commitee of the 
 Brazilian National Meeting on Theoretical and Computational Physics, 
April 6 to 9, 2003, Brasilia-DF, Brazil.


\begin{thebibliography}{88}


\bibitem{W} E. Witten, {\it Nucl. Phys.} {\bf B185},
513 (1981); 
R. de Lima Rodrigues, ``The Quantum Mechanics SUSY
Algebra: an Introductory Review,'' {\it Monograph CBPF}
MO-03/01, e-print hep-th/0205017 and references therein.

\bibitem{semenove91} Yu V. Ralchenko and V. V. Semenov,
{\it J. Phys. A,} {\bf 24}, 
L1305  (1991).

\bibitem{ss2} R. D. Amado, F. Cannata and J. P. Dedonder, 
{\it Int. J. Mod. Phys.} {\bf A5}, 3401 (1990);
F. Cannata and M. V. Ioffe, {\it Phys. Lett.} {\bf B278}, 399 (1992);
F. Cannata and M. V. Ioffe, {\it J. Phys. A,} {\bf 26}, 
L89  (1993);
X.-Y. Wang, B.-C. Xu and P. L. Taylor, {\it Phys. Lett.}  
{\bf A 137}, 30 (1993);
A. Andrianov, F. Cannata, M. V. Ioffe and D. N. Nishnianidze 
{\it J. Phys. A,} {\bf 30}, 5037 (1997);
T. K. Das and B. Chakrabari, {\it J. Phys. A,}  
{\bf 32} 2387 (1999).

\bibitem{Ta93} T. Fukui, {\it Phys. Lett.} {\bf A178}, 1 (1993).

\bibitem{ger03} G. S. Dias, E. L. Gra\c{c}a and R. de Lima Rodrigues,
"Stability equation and two-component eigenmode for damain walls in a
scalar potential model," hep-th/0205195.

\bibitem{RVV} L. Vestergaard Hau, J. A. Golovchenko and Michael M. Burns,
{\it Phys. Rev. Lett.}  {\bf 74}, 3138 (1995);
R. de Lima Rodrigues,
V. B. Bezerra and A. N. Vaidya,
{\it Phys. Lett.} {\bf A287}, 45 (2001), hep-th/0201208.

\end{thebibliography}
\end{document}